\documentstyle[12pt]{article}

\clearpage
\begin{document}

\hfill SOGANG-HEP 238/98

\hfill October 2, 1998

\vspace{0.5cm}

\begin{center}
{\large \bf Two Different Gauge Invariant Models \\in Lagrangian Approach}
\end{center}

\vspace{0.5cm}

\begin{center}
Seung-Kook Kim$^{*}$, Yong-Wan Kim$^{\dagger}$,
and Young-Jai Park$^{\ddagger}$ \\
\end{center}

\begin{center}
{\it \ $^{*}$ Department of Physics, Seonam University,
Namwon, Chonbuk 590-170, Korea}\\
and\\
{\it \ $^{\dagger~\ddagger}$ Department of Physics 
and Basic Science Research Institute \\
Sogang University, C.P.O. Box 1142, Seoul 100-611, Korea} 
\end{center}

\vspace{1cm}

\begin{center}
{\bf ABSTRACT}
\end{center}

We show how to systematically derive the complete set of
the gauge transformations of different types of the gauge invariant models,
which are the chiral Schwinger and CP$^1$ with Chern-Simons term, 
in the Lagrangian Formalism.

\vspace{1cm}

PACS number : 11.10.Ef, 11.15.Tk, 11.15.-q \\

Keywords: Lagrangain approach; Gauge Invariant; Chiral Schwinger Model;
CP$^1$ Model; Hamiltonian embedding.

\vspace{4cm}

\noindent $*$ e--mail address: skandjh@chollian.net\\
$\dagger$ e--mail address: ywkim@physics3.sogang.ac.kr\\
$\ddagger$ e--mail address: yjpark@ccs.sogang.ac.kr

\newpage
\begin{center}
\section{\bf Introduction}
\end{center}

The Hamiltonian embedding [1-6] of constrained systems has the drawback 
of not necessarily leading to a manifestly Lorentz covariant partition 
function. This problem could be avoided in the Lagrangian field--anti-field 
approach [7],
which is based on an analysis of local symmetries of a Lagrangian. 
The establishment of the full, irreducible set of local symmetries of 
a Lagrangian thus plays a fundamental role in this formalism.
By the way, these symmetries are often put, by hand, 
in the action while constructing the Lagrangian, and sometimes 
it is found by direct observation or trial and error.
Moreover, it may be, that for some complicated Lagrangians 
the full local symmetries can not be seen directly. 
Their systematic and exhaustive determinations of gauge symmetries 
structure thus constitute an integral part of the field--anti-field 
quantization program without the use of Dirac's Hamiltonian construction 
of the corresponding generators.

On the other hand, Batalin {\it et al.} (BFT) Hamiltonian method 
[1] has been applied to second class constrained systems [2,3], 
which yield the strongly involutive first class
constraint algebra in an extended phase space.
Recently, we have quantized other interesting models including the Proca 
models by using our improved BFT formalism [4--6].

In this paper, we will consider Lagrangian approach 
of the different types of gauge invariant systems, 
which have different constraint structure in the Lagrangian sense. 
According to the classifications of constraints in Ref. [8],
the A-type constraints are defined by functions 
without having velocities, while  
the B-type ones are a set of functions of velocity and coordinates.
With these types of classifications, we will show that
their ``identically'' vanishing parts of the successive evolutions
for the stability of the constraints systematically 
generate gauge symmetry of system.
Recently, Shirzad [9] tried to apply this Lagrangian formulation
to the chiral Schwinger model (CSM) as well as Schwinger model. 
However, he could not obtain the 
complete set of gauge transformations because he used the anomalous CSM,
which is a gauge non-invariant second class constraint system.
Thus, we will briefly recapitulate the BFT Hamiltonian embedding [4,10]
of the gauge non-invarinat CSM with $a > 1$ [11] in section 2,
in order to show how one can systematically construct the gauge invariant CSM. 
Then, we explicitly show how
to derive the exact form of gauge transformation 
of the first class CSM model, which is the type B, 
making use of the iteratated Lagrangian equations of motion [9,12].
In section 3, we also derive that the well--known form of gauge 
transformation of the gauge invariant CP$^1$ model with 
the Chern-Simons (CS) term [13--16], 
which is the case of being mixed with the type A and B constraints.
Our conclusions are given in section 4.

\vspace{1cm}
\begin{center}
{\section {\bf Chiral Schwinger Model with Wess-Zumino (WZ) term}}
\end{center}

We first briefly recapitulate our previous BFT Hamiltonian embedding [4,10] 
of the bosonized CSM model in the case of $a>1$, 
whose dynamics are given by
\begin{equation}
\label{1}
    S_{CSM} ~=~ \int d^2x~\left[
                          -\frac{1}{4}F_{\mu \nu}F^{\mu \nu}
                          +\frac{1}{2}\partial_{\mu}\phi\partial^{\mu}\phi
                          +eA_{\nu}(\eta^{\mu \nu}
                          -\epsilon^{\mu \nu})\partial_{\mu}\phi
              +\frac{1}{2}ae^{2}A_{\mu}A^{\mu}~\right],
\end{equation}
where $\eta^{\mu\nu}=\mbox{diag(1,-1)}$, $\epsilon^{01}=1$, and
$a$ is a regularization ambiguity [11].
The canonical momenta are given by
$\pi_0=0, \pi_1=F^{01}$, and $\pi_\phi=\dot{\phi}+e(A^0+A^1)$.
One then finds one primary [17] constraint
$\Omega_1 \equiv \pi_0 \approx 0$
and one secondary constraint
$\Omega_2 \equiv \partial^1 \pi_1 + e\pi_\phi + e\partial_1 \phi
                       - e^2 A^1 + (a-1)e^2 A^0 \approx 0$,
which is obtained by requiring the consistency of 
the primary constraint $\Omega_1$ with time evolution.
These constraints fully form the second class:
\begin{eqnarray}
\label{2}
        \Delta_{ij}(x,y)
                       &:=&
                \{ \Omega_{i}(x), \Omega_{j}(y) \} \nonumber\\
            &=& - e^2 (a-1)\epsilon_{ij}\delta(x-y), 
\end{eqnarray}
and the canonical Hamiltonian $H_c$ is given by
\begin{eqnarray}
\label{3}
    H_c &=& \int\!dx~\left[~
            \frac{1}{2}(\pi_1)^2 + \frac{1}{2}(\pi_{\phi})^2
             + \frac{1}{2}(\partial_{1}\phi)^2 - e(\pi_{\phi}
             + \partial_{1}\phi)(A^0 + A^1) \right. \nonumber \\
        &&~~~~~~ \left. - A^0\partial^1\pi_1
               - \frac{1}{2}ae^{2} \{ (A^0)^2 - (A^1)^2 \}
               + \frac{1}{2}e^{2}(A^0 + A^1)^2~\right].
\end{eqnarray}

We now introduce auxiliary fields $\Phi^{i}$
in order to convert the second class constraint $\Omega_{i}$ into
first class ones in an extended phase space. 
Following the BFT Hamiltonian formalism [1,4--6,10], 
we require these fields to satisfy
\begin{eqnarray}
\label{4}
   \{ \Phi^i(x), \Phi^j(y) \} &=& \omega^{ij}(x,y)=\epsilon^{ij}\delta(x-y),\nonumber\\
   \{ F, \Phi^{i} \} &=& 0,
\end{eqnarray}
where $F$ denotes the original variables, 
$(A^{\mu},\pi_{\mu},\phi, \pi_{\phi} )$, collectively.
Strongly involutive constraints
$\widetilde{\Omega}_{i}$  satisfying the requirement of 
$\{\widetilde{\Omega}_{i}, \widetilde{\Omega}_{j} \}=0$
as well as the boundary conditions,
$\widetilde{\Omega}_i \mid_{\Phi^i = 0} = \Omega_i$,
can be obtained in power series as 
$\tilde\Omega_i=\sum_n\Omega^{(n)}_i$, where $\Omega^{(n)}_i$ is proportional
to $(\Phi^j)^n$.
The first order correction terms in the infinite series are simply given
by $\Omega^{(1)}_i=\int dy \ X_{ij}(x,y)\Phi^j(y)$. 
Then, the strongly involutive relation of $\tilde\Omega_i$ provide the following
condition
\begin{equation}
\label{411}
\Delta_{ij}(x,y)+\int du dv \ X_{ik}(x,u)\omega^{k\ell}(u,v)X_{j\ell}(v,y)=0,
\end{equation}
from which we get a solution of $X_{ij}=e\sqrt{a-1}\delta_{ij}\delta(x-y)$
in case of choosing $\omega^{ij}$ as in Eq. (\ref{4}).
Now, making use of $\omega^{ij}$ and $X_{ij}$, we can easily obtain the strongly
involutive constraints as
$\widetilde{ \Omega}_{i}=\Omega_{i} +e \sqrt{a-1} \Phi^{i}$.

On the other hand, corresponding to the original variables
$F$, strongly involutive BFT variables
$\widetilde{F} \equiv (\widetilde{A}^{\mu},\widetilde{\pi}_{\mu},
\widetilde{\phi}, \widetilde{\pi}_{\phi} )$ such as
$\{ \widetilde{\Omega}_{i}, \widetilde{F} \} =0$,
are given by
\begin{eqnarray}
\label{5}
\widetilde{A}^{\mu} &=& (A^{0} + \frac{1}{e \sqrt{a-1}} \Phi^{2},
      A^{1} - \frac{1}{e \sqrt{a-1}} \partial_1 \Phi^{1} ),  \nonumber \\
\widetilde{\pi}_{\mu} &=& (\pi_{0}+e \sqrt{a-1} \Phi^{1}, 
                         \pi_{1}+\frac{e}{\sqrt{a-1}} \Phi^{1} ), \nonumber \\
\widetilde{\phi} &=& \phi - \frac{1}{ \sqrt{a-1}} \Phi^{1}, \nonumber \\
\widetilde{\pi}_\phi &=& \pi_{\phi} - \frac{1}{ \sqrt{a-1}} \partial_1 \Phi^{1}.
\end{eqnarray}
Using these BFT fields,
we can find the desired first class Hamiltonian $\widetilde{H}$
from the canonical Hamiltonian $H_{c}$ as 
\begin{eqnarray}
\label{6}
   \widetilde{H}(A^{\mu}, \pi_{\nu},\phi, \pi_{\phi}; \Phi^{i} )
      \!\!&=&\!\! H_{c}(\widetilde{A}^{\mu}, \widetilde{\pi}_{\nu}
      , \widetilde{\phi},\widetilde{ \pi}_{\phi}) \nonumber \\
  \!\!&=&\!\! H_{c}(A^{\mu}, \pi_{\nu},\phi, \pi_{\phi})  
       +\int\! dx
      \left[
         \frac{1}{2}(\partial_1 \Phi^1)^2
         + \frac{e^2}{2(a-1)}(\Phi^1)^2 
         + \frac{1}{2}(\Phi^2)^2  \right. \nonumber \\
&& \left. 
      - \frac{1}{e\sqrt{a-1}}  [e^2 \pi_1 - e^2(a-1) \partial_1 A^1 ]
          \Phi^1 - \frac{1}{e\sqrt{a-1}}\Phi^2\widetilde{\Omega}_2 
      \right],
\end{eqnarray}
which, by construction, is automatically strongly involutive, {\it i.e.},
$\{ \widetilde{\Omega}_{i}, \widetilde{H} \} =0$.

It seems appropriate to comment on generators of local symmetry 
transformation in the Hamiltonian formulation, 
which are fully given by the first class constraints. 
Defining the generators by
\begin{equation}
\label{7}
G:=\sum^2_{\alpha=1}\int d^2x~(-1)^{\alpha+1}\epsilon^\alpha(x)
\widetilde\Omega_\alpha(x),
\end{equation}
we have $(\delta A=\{A,G\})$
\begin{eqnarray}
\label{8}
&&\delta A^0=\epsilon^1,~~~~~~\delta \pi_0=(a-1)e^2\epsilon^2, \nonumber\\
&&\delta A^1=-\partial_1\epsilon^2,~~~~\delta \pi_1=-e^2\epsilon^2,\nonumber\\
&&\delta\phi=-e\epsilon^2,~~~~~~\delta \pi_\phi=e\partial^1\epsilon^2, 
\nonumber\\
&&\delta\theta=-e\epsilon^2,~~~~~~\delta \pi_\theta=-(a-1)e\epsilon^1.
\end{eqnarray}
Without loss of any generality, we have inserted $(-1)^{\alpha+1}$ factor 
in Eq. (\ref{7}) in order to make the gauge transformation as usual, 
and also identified the new variables
$\Phi^i$ as a canonically conjugate pair,
$\Phi^i=(\sqrt{a-1}\theta, \frac{1}{\sqrt{a-1}}\pi_\theta)$, 
satisfying Eq. (\ref{4}).

Now, we consider the partition function of the model
in order to present the Lagrangian corresponding to $\widetilde{H}$.
The starting phase space partition function is given by the 
Faddeev--Popov formula [18] as follows
\begin{equation}
\label{9}
Z=  \int  {\cal D} A^\mu
          {\cal D} \pi_\mu
          {\cal D} \phi
          {\cal D} \pi_{\phi}
          {\cal D} \theta
          {\cal D} \pi_\theta
               \prod_{i,j = 1}^{2} \delta(\widetilde{\Omega}_i)
                           \delta(\Gamma_j)
                \mbox{det} \mid \{ \widetilde{\Omega}_i, \Gamma_j \} \mid
                e^{iS'},
\end{equation}
where
\begin{equation}
\label{10}
S'  =  \int d^2x \left(
               \pi_\mu {\dot A}^\mu +\pi_{\phi} {\dot \phi} 
         + \pi_\theta {\dot \theta} - \widetilde {\cal H}'
                \right),
\end{equation}
and we have used an equivalent first class Hamiltonian
by adding a term proportional to the first class constraint
$\widetilde{\Omega}_2$:
\begin{equation}
\label{11}
\widetilde{H}' = \widetilde H 
        + \int d x \frac{1}{(a-1)e} \pi_{\theta} \widetilde{\Omega}_2.
\end{equation}
Note that the gauge fixing functions $\Gamma_j$ are chosen so that the
determinant occurring in the functional measure is nonvanishing.
The  $\pi_0$ integration is trivially performed by exploiting
the delta function $\delta(\widetilde{\Omega}_1)=\delta[\pi_0 + (a-1)e\theta]$.
After exponentiating the remaining delta function $
\delta(\widetilde{\Omega}_2) =
    \delta[ \partial^1 \pi_1 + e\pi_\phi + e\partial_1 \phi
            - e^2 A^1 + (a-1)e^2 A^0 + e\pi_\theta ] $ 
in terms of a
Fourier variable $\xi$ as $\delta(\widetilde{\Omega}_2)=
\int{\cal D}\xi e^{-i\int d^2x\xi\tilde{\Omega}_2}$, 
transforming $A^0 \to A^0 + \xi$,
and integrating the momentum variables $\pi_\phi$,
$\pi_1$, and $\pi_\theta$, 
one could obtain the gauge invariant action up to a total divergence
as follows
\begin{eqnarray}
\label{12}
    S &=& S_{CSM} + S_{WZ} ~;            \\
\label{13}
    S_{WZ} &=&  \int d^2x \left[
                \frac{1}{2}(a-1)\partial_\mu \theta \partial^\mu \theta
                + e \partial_{\mu}\theta
                   \{ (a-1)\eta^{\mu\nu} + \epsilon^{\mu\nu} \}
                     A_\nu  \right],
\end{eqnarray}
where $S_{WZ}$ is the well-known WZ term, which serves to cancel the
gauge anomaly.
The corresponding measure now reads
\begin{equation}
\label{14}
[{\cal D} \mu] = {\cal D} A^\mu
         {\cal D} \phi
                 {\cal D} \theta
                 {\cal D} \xi
                 \prod^2_{\beta = 1}
           \delta\left(\Gamma_{\beta}[A^0 + \xi, A^1, \phi, \theta]\right)
        \det \mid \{\widetilde{\Omega}_{\alpha}, \Gamma_{\beta}\} \mid.
\end{equation}

Now, we are ready to apply the Lagrangian approach [9,12]
to the gauge invariant action (\ref{12}).
The equations of motion are of the form
\begin{equation}
\label{15}
L^{(0)}_i(x):=\int dy \left[W^{(0)}_{ij}(x,y)\ddot\varphi^j(y)
+\alpha^{(0)}_i(y)\delta(x-y)\right]=0, \ i=1,...,4,
\end{equation}
where
$W^{(0)}_{ij}(x,y)$ is the Hessian matrix
\begin{eqnarray}
\label{16}
W^{(0)}_{ij}(x,y)&:=&
       \frac{\delta^2{\cal L}}{\delta\dot\varphi^i(x)\delta\dot
       \varphi^j(y)}\nonumber\\
&=& 
\left(\begin{array}{cccc}
       0 & 0 & 0 & 0  \\
       0 & 1 & 0 & 0  \\
       0 & 0 & 1 & 0  \\
       0 & 0 & 0 & a-1  
       \end{array} 
       \right)\delta(x-y)
=\widetilde{W}^{(0)}_{ij}\delta(x-y),\\
\label{17}
(\varphi^i)^{\rm T}(x)
&:=&(A^0,A^1,\phi,\theta)(x),\\
\label{18}
(\alpha^{(0)}_i)^{\rm T}(x)&:=&\int dy\left[
\frac{\partial^2{\cal L}}{\partial\varphi^j(y)\partial\dot\varphi^i(x)}
\partial\dot\varphi^j(y)\right]-\frac{\partial {\cal L}}{\partial\varphi^i(x)}
\nonumber\\
&=&(\alpha_{A^0},\alpha_{A^1},\alpha_\phi,\alpha_\theta)(x)
\end{eqnarray}
with
\begin{eqnarray}
\label{19}
&&\alpha_{A^0}=\partial_1(\dot{A}^1+\partial_1A^0)-e\dot{\phi}-e\partial_1
\phi-ae^2A^0-(a-1)e\dot{\theta}+e\partial_1\theta,
\nonumber\\
&&\alpha_{A^1}=\partial_1\dot{A}^0-e\dot{\phi}-e\partial_1\phi+ae^2A^1
+e\dot{\theta}-(a-1)e\partial_1\theta,
\nonumber\\
&&\alpha_\phi=e\dot{A}^0+e\dot{A}^1-\partial^2_1\phi+e\partial_1
A^0+e\partial_1A^1, \nonumber\\
&&\alpha_\theta=(a-1)e\dot{A^0}-e\dot{A}^1-(a-1)\partial^2_1\theta
-e\partial_1A^0+(a-1)e\partial_1A^1.
\end{eqnarray}
The Hessian matrix (\ref{16}) is of rank three.
Hence there exists a ``zeroth generation'' null eigenvector
$\lambda^{(0)}(x,y)$ satisfying
\begin{equation}
\label{20}
\int dy\ \lambda^{(0)}_i(x,y)\ W^{(0)}_{ij}(y,z)=0.
\end{equation}
For simplicity, let us normalize it to have components
\begin{equation}
\label{21}
\lambda^{(0)}(x,y)=(1,0,0,0)\delta(x-y).
\end{equation}
Correspondingly we have a``zeroth generation" constraint
\begin{equation}
\label{22}
\Omega^{(0)}_1(x)=\int dy \ \lambda^{(0)}_i(x,y) L^{(0)}_i(y)=L^{(0)}_1(x)
=\alpha_{A^0}=0,
\end{equation}
in the Lagrangian sense.

Similar to the time stability condition of constraints in the Hamiltonian
formalism, 
we now require the primary Lagrange constraint (\ref{22})
to be independent of time. We thus need to add to the equations
of motion (\ref{15}) through the equation $\dot{\Omega}^{(0)}_1=0$. 
Then, the resulting set of five equations may be summarized
in the form of the set of ``first generation'' equations,
$L^{(1)}_{i_1}(x)=0,\ i_1=1,...,5$, with
\begin{equation}
\label{23}
L_{i_1}^{(1)}(x):=\left\{\begin{array}{ll}
L_i^{(0)},\ i=1,...4,\\
\frac{d}{dt}(\lambda^{(0)}_iL^{(0)}_i).
\end{array}\right.
\end{equation}
$L_{i_1}^{(1)}(x)$  can be written in the general form
\begin{equation}
\label{24}
L_{i_1}^{(1)}(x):
=\int dy \left[W_{i_1j}^{(1)}(x,y)\ddot\varphi^j(y)
            +\alpha^{(1)}_{i_1}(y)\delta(x-y)\right]=0,
\end{equation}
where
\begin{equation}
\label{25}
W^{(1)}_{i_1j}(x,y)=\left(\begin{array}{cc}
\\
\widetilde{W}^{(0)}_{ij}\\
\\
\hline\\
\begin{array}{ccccc}
0&\partial^x_1&-e&-(a-1)e
\end{array}
\end{array}
\right)\delta(x-y),
\end{equation}
and
\begin{equation}
\label{26}
(\alpha^{(1)}_{i_1})^{\rm T}(x)=( (\alpha^{(0)}_i)^{\rm T}, \alpha^{(1)}_5)(x)
\end{equation}
with
\begin{eqnarray}
\label{27}
\alpha^{(1)}_5=\partial^2_1\dot{A}^0-ae^2\dot{A}^0+e\partial_1
\dot{\phi}-e\partial_1\dot{\theta}.
\end{eqnarray}

Next, let us repeat the previous analysis 
taking Eq. (\ref{24}) as a starting point, 
and looking for solutions of a first generation null eigenvector of
\begin{equation}
\label{28}
\int dy \ \lambda^{(1)}_{i_1}(x,y)W^{(1)}_{i_1j}(y,z)=0.
\end{equation}
Since $W^{(1)}_{i_1j}(x,y)$ is still of rank three, there exists 
a null eigenvector, $\lambda^{(1)}(x,y)$, with the previous eigenvector
extended as $\lambda^{(0)}(x,y)=(1,0,0,0,0)\delta(x-y)$. 
This $\lambda^{(1)}(x,y)$ is of the form 
$(0,-\partial_1^x,e,e,1)v(x)\delta(x-y)$. 
We could thus choose it as
\begin{equation}
\label{29}
\lambda^{(1)}_{i_1}(x,y)=(0,-\partial_1^x,e,e,1)\delta(x-y).
\end{equation}
The associated constraint is found to vanish ``identically'':
\begin{equation}
\label{30}
\Omega^{(1)}_2(x)=\int dy \ \lambda^{(1)}_{i_1}(x,y)L^{(1)}_{i_1}(y)
=-\partial_1\alpha^{(1)}_2 + e\alpha^{(1)}_3+ e\alpha^{(1)}_4+\alpha^{(1)}_5=0.
\end{equation}
Therefore, the algorithm ends at this stage.

The local symmetries of the action (\ref{12}) 
are encoded in the identity (\ref{30}). 
Recalling (\ref{15}) and (\ref{24})
we see that the identity (\ref{30}) can be rewritten as follows
\begin{eqnarray}
\label{31}
\Omega^{(1)}_2(x)=-\partial_1L^{(0)}_2+eL^{(0)}_3+eL^{(0)}_4
+\frac{d}{dt}L^{(0)}_1\equiv0.
\end{eqnarray}
This result is a special case of a general theorem stating
[8,9] that the identities $\Omega^{(l)}_k\equiv 0$ 
can always be written in the form
\begin{equation}
\label{32}
\Omega^{(l)}_k:=\sum_{s=0} \int dy \left((-1)^{s+1}
                \frac{d^s}{dt^s}\phi^{i(s)}_k(x,y)L^{(0)}_i(y)\right).
\end{equation}
Then, for the gauge invariant CSM, we have the following relations
\begin{eqnarray}
\label{33}
&&\phi^{2(0)}_2(x,y)=\partial^x_1\delta(x-y),\nonumber\\
&&\phi^{3(0)}_2(x,y)=-e\delta(x-y),\nonumber\\
&&\phi^{4(0)}_2(x,y)=e\delta(x-y),\nonumber\\
&&\phi^{1(1)}_2(x,y)=-\delta(x-y),
\end{eqnarray}
while all the others are vanishing.

On the other hand, it follows from general considerations [8,9]
that the action (\ref{12}) is invariant under the transformation
\begin{equation}
\label{34}
\delta\varphi^i(y):=\sum_{k}\int dx \ \left(\Lambda_k(x)\phi^{i(0)}_k(x,y)
                    + \dot{\Lambda}_k(x)\phi^{i(1)}_k(x,y)\right).
\end{equation}
For the CSM case this corresponds to the transformations
\begin{eqnarray}
\label{35}
&&\delta A^\mu(x)=\partial^\mu\Lambda_2,\nonumber\\
&&\delta \phi(x)=-e\Lambda_2, \nonumber\\
&&\delta \theta(x)=-e\Lambda_2.
\end{eqnarray}
These are the set of symmetry transformations which is identical with 
the previous result (\ref{8}) of the extended Hamiltonian formalism,
when we set $\epsilon^1=\partial^0\epsilon^2$ and $\epsilon^2=\Lambda_2$, 
similar to the Maxwell case [19].
As results, we have systematically derived the set of symmetry 
transformations starting from the Lagrangian of the gauge invariant CSM.

\vspace{1cm}
\begin{center}
{\section{\bf CP$^1$ Model with the CS term}}
\end{center}

The CP$^1$ model with the CS term [13,14],
which is an archetype example of field theory
and the constrained system of a mixed Type A and B,
was considered by Polyakov who found Bose-Fermi statistics
transmutation [20] in the model.
Han [15] has analyzed this CP$^1$ model by using
the Dirac formalism together with the first-order Lagrangian method.
We has recently analyzed  the CP$^1$ model with the CS term
by fully using the symplectic formalism [16].

Our starting Lagrangian for the gauge-invariant CP$^1$ model with the
CS term [13--16] to analyze in the Lagrangian approach 
is given by
\begin{equation}
\label{36}
{\cal L} = \frac{\kappa}{2\pi} \epsilon^{\mu \nu \rho}
            A_{\mu} \partial_{\nu} A_{\rho}
             + (\partial_{\mu} + {\it i} A_{\mu}) z^*_a
             (\partial^{\mu} - {\it i}
             A^{\mu}) z_a ;~~~ a = 1,2
\end{equation}
with the CP$^1$ constraint
\begin{equation}
\label{37}
\Omega = |z_a|^2 - 1 = 0,
\end{equation}
where the convention is $ \eta^{\mu \nu} = {\rm diag} (1, -1, -1)$ 
and $\epsilon^{012}=+1$.

The equations of motion from the Lagrangian (\ref{36}) 
can be written of the form
\begin{equation}
\label{38}
L_i^{(0)} (x) := \int d^2 x 
\left [ W_{ij}^{(0)} (x,y) \ddot \varphi^j (y) + \alpha_i^{(0)} (y) 
\delta^2 (x-y) \right ] =0, ~~~i = 1, \cdots , 7 
\end{equation}
where $W_{ij}^{(0)} (x,y)$ is
\begin{eqnarray}
\label{39}
W^{(0)}_{ij}(x,y) &=& 
\left(\begin{array}{ccccccc}
       0 & 0 & 0 & 0 & 0 & 0 & 0 \\
       0 & 0 & 0 & 0 & 0 & 0 & 0 \\
       0 & 0 & 0 & 0 & 0 & 0 & 0 \\
       0 & 0 & 0 & 0 & 0 & 1 & 0 \\
        0 & 0 & 0 & 0 & 0 & 0 & 1 \\
        0 & 0 & 0 & 1 & 0 & 0 & 0 \\
        0 & 0 & 0 & 0 & 1 & 0 & 0
       \end{array} 
       \right) \delta^2 (x-y)
=\widetilde{W}^{(0)}_{ij} \delta^2 (x-y),\\
\label{40}
(\varphi^i)^{\rm T}(x)
&=&(A^0,A^1,A^2,z_1,z_2,z_1^*,z_2^*)(x),\\
\label{41}
(\alpha^{(0)}_i)^{\rm T}(x) 
&=&(\alpha_{A^0},\alpha_{A^1},\alpha_{A^2},\alpha_{z_1},\alpha_{z_2}, \alpha_{z_1^*}, \alpha_{z_2^*} )(x)
\end{eqnarray}
with
\begin{eqnarray}
\label{42}
&&\alpha_{A^0} = - \frac{\kappa}{\pi} \epsilon_{mn} \partial^m A^n 
- i (z_a^* \dot{z}_a - \dot{z}_a^*z_a) 
- 2 |z_a|^2 A^0,\nonumber \\
&&\alpha_{A^m} = - \frac{\kappa}{\pi} \epsilon_{nm} \dot A^n 
+ \frac{\kappa}{\pi} \epsilon_{nm}\partial^n A^0 + i (z_a^* \partial^m z_a 
- z_a \partial^m z_a^*) + 2 |z_a|^2 A^m, \nonumber \\
&&\alpha_{z_a} = i z_a^* \dot A^0  + 2i A^\mu \partial_\mu z_a^* 
+ i z^*_a \partial_m A^m - z_a^* A_\mu A^\mu 
+ \partial_m \partial^m z_a^*, \nonumber \\
&&\alpha_{z_a^*} = -i z_a \dot A^0  - 2i A^\mu \partial_\mu z_a 
- i z_a \partial_m A^m - z_a A_\mu A^\mu +\partial_m\partial^m z_a. 
\end{eqnarray}
Note that here $m=1,2$.
The Hessian matrix (\ref{39}) is of rank four. 
Hence there exist three ``zeroth generation" null eigenvectors 
$\lambda^{(0) A} (x,y)$ satisfying
\begin{equation}
\label{43}
\int d^2y\ \lambda^{(0)A}_i(x,y)\ W^{(0)}_{ij}(y,z)=0 , ~~~~~A=1,2,3.
\end{equation}
We choose them to have components
\begin{eqnarray}
\label{44}
\lambda_i^{(0)1} (x,y) &=& (1,0,0,0,0,0,0 ) \delta^2 (x-y), \nonumber \\
\lambda_i^{(0)2} (x,y) &=& (0,1,0,0,0,0,0 ) \delta^2 (x-y), \nonumber \\
\lambda_i^{(0)3} (x,y) &=& (0,0,1,0,0,0,0 ) \delta^2 (x-y). 
\end{eqnarray}
Correspondingly we have the ``zeroth generation" constraints, which are B-type
\begin{equation}
\label{45}
\Omega_k^{(0)} = \alpha_k = 0 , ~~~~ k=1,2,3.
\end{equation}
On the other hand, in this CP$^1$ case 
we have also one more constraint, {\it i.e.}, 
CP$^1$ constraint (\ref{37}) which is A-type.
Since the time derivative of $\Omega$ and 
${d\Omega\over dt} = z_a^* \dot z_a + z_a \dot z_a^*$
are independent of $\Omega_k$, we can obtain the following constraints
\begin{eqnarray}
\label{46}
\Omega_k^{(0)} &=& \alpha_k, \nonumber \\
\Omega_4^{(0)} &=& \frac{d\Omega}{dt} = z_a^* \dot z_a + z_a \dot z_a^*,
\end{eqnarray}
on the first stage of iteration.

Using the consistency condition for the constraints with time, 
we need to add the equation $\dot \Omega_{k'}^{(0)} = 0 ,~(k'= 1,\cdots,4)$
to the equation of motion (\ref{38}). 
Hence we have the set of ``first generation" equations, 
$L_{i_1}^{(1)} (x) = 0 ,~ i_1 = 1, \cdots , 11$, as follows
\begin{equation}
\label{47}
L_{i_1}^{(1)}(x):=\left\{\begin{array}{lll}
L_i^{(0)},~~~\ i_1=1,...7,\\
\frac{d}{dt}(\lambda^{(0) A}_iL^{(0)}_i)~~~i_1 = 7 +A ,~A=1,2,3, \\
\frac{d}{dt} ( \Omega_4^{(0)} ) ~~~i_1 = 11.
\end{array}\right.
\end{equation}
$L_{i_1}^{(1)}(x)$ can be written in the general form
\begin{equation}
\label{48}
L_{i_1}^{(1)}(x):
=\int d^2y \left[W_{i_1j}^{(1)}(x,y)\ddot\varphi^j(y)
            +\alpha^{(1)}_{i_1}(y)\delta^2 (x-y)\right]=0,
 ~~~i_1 = 1, \cdots, 11,
\end{equation}
where
\begin{equation}
\label{49}
W^{(1)}_{i_1j}(x,y)=\left(\begin{array}{ccccc}
\\
\widetilde{W}^{(0)}_{ij}\\
\\
\hline\\
\begin{array}{ccccccc}
0 & 0 & 0 & -iz_1^* & -iz_2^* & iz_1 & iz_2 \\
0 & 0 & \frac{\kappa}{\pi} & 0 & 0 & 0 & 0  \\
0 & - \frac{\kappa}{\pi} & 0 & 0 & 0 & 0 & 0 \\
0 & 0 & 0 & z_1^* & z_2^* & z_1 & z_2
\end{array}
\end{array}
\right)\delta^2(x-y),
\end{equation}
and
\begin{equation}
\label{50}
( \alpha_{i_1}^{(1)} )^{\rm T} (x) 
= ( (\alpha_i^{(0)})^{\rm T}, \alpha_8^{(1)}, 
\alpha_9^{(1)}, \alpha_{10}^{(1)}, \alpha_{11}^{(1)} ) (x),
\end{equation}
with
\begin{eqnarray}
\label{51}
\alpha_8^{(1)} &=& -2|z_a|^2 \dot A^0 
- \frac{\kappa}{\pi} \epsilon_{m n} \partial^m \dot A^n 
- 2 A^0 \frac{d}{dt} |z_a|^2, \nonumber \\
\alpha_9^{(1)} &=& - \frac{\kappa}{\pi} \partial^2 \dot A^0 
+ 2 |z_a|^2 \dot A^1 + 2 A^1 \frac{d}{dt} |z_a|^2 
+ i \frac{d}{dt}(z_a^* \partial^1 z_a - z_a \partial^1 z_a^*) \nonumber \\
\alpha_{10}^{(1)} &=& \frac{\kappa}{\pi} \partial^1 \dot A^0 
+ 2 |z_a|^2 \dot A^2 + 2 A^2 \frac{d}{dt}|z_a|^2
+ i \frac{d}{dt}( z_a^* \partial^2  z_a - z_a \partial^2 \dot z_a^*), \nonumber \\
\alpha_{11}^{(1)} &=& 2 \dot z_a \dot z_a^*,
\end{eqnarray}
respectively.

Since $W_{ij}^{(1)} (x,y)$ is of rank six, 
there exist two ``first generation" null eigenvectors 
$\lambda^{(1) A}(x,y),~A=1,2$
with the previous three null eigenvectors extended as in the section 2.
Similarly using the Eq. (\ref{28}), 
these null eigenvectors are explicitly given by
\begin{eqnarray}
\label{52}
\lambda_{i_1}^{(1)1}(x,y) &=& 
( 0, 0, 0, -2iz_1, -2iz_2, 0, 0, 1, 0, 0, i)\delta^2(x-y), 
\nonumber \\  
\lambda_{i_1}^{(1) 2}(x,y) 
&=& ( 0, 0, 0 , 0, 0, -2iz_1^* , -2iz_2^*, -1, 0, 0, i)\delta^2(x-y).
\end{eqnarray} 
Associated with this eigenvectors we have new constraints
\begin{eqnarray}
\label{53}
\Omega_1^{(1)} &=& 2i\dot z_a \dot z_a^* + 2A^0 (z_a \dot z_a^* - z_a^* \dot z_a) 
- \frac{\kappa}{\pi} \epsilon_{m n} \partial^m \dot A^n 
+ 4 z_a A^m \partial_m z_a^* + 2 |z_a|^2 \partial_m A^m \nonumber\\
~~~~~&&+ 2i|z_a|^2 A_\mu A^\mu  - 2i z_a \partial_m \partial^m z_a^*, \nonumber \\
\Omega_2^{(1)} &=& 2i\dot z_a \dot z_a^* + 2A^0 (z_a \dot z_a^* -  z_a^* \dot z_a) 
+ \frac{\kappa}{\pi} \epsilon_{m n} \partial^m \dot A^n 
- 4 z_a^* A^m \partial_m z_a - 2 |z_a|^2 \partial_m A^m  \nonumber \\ 
~~~~~&& + 2i |z_a|^2 A_\mu A^\mu  - 2i z_a^* \partial_m \partial^m z_a.
\end{eqnarray}

We now repeat the above procedure 
by using the consistency condition for the constraints with time, 
and obtain the ``second generation" equation,
$L_{i_2}^{(2)} (x)=0,~i_2 = 1, \cdots, 13$, with
\begin{equation}
\label{54}
L_{i_2}^{(2)}(x):=\left\{\begin{array}{llll}
L_i^{(0)},~~~\ i_1=1,...7,\\
\frac{d}{dt}(\lambda^{(0) A}_iL^{(0)}_i),~~~i_2 = 7 +A ,~A=1,2,3, \\
\frac{d}{dt} ( \Omega_4^{(0)} ), ~~~i_2 = 11, \\
\frac{d}{dt} ( \lambda_{i_1}^{(1) B} L_{i_1}^{(1)}),~~~i_2 = 11+B , ~B=1,2.
\end{array}\right.
\end{equation}
The resulting complete set of equation is of the form
\begin{equation}
\label{55}
L_{i_2}^{(2)}(x):
=\int d^2y \left[W_{i_2j}^{(2)}(x,y)\ddot\varphi^j(y)
            +\alpha^{(2)}_{i_2}(y)\delta^2 (x-y)\right]=0,~~~i_2 = 1, 
\cdots, 13,
\end{equation}
where
\begin{equation}
\label{56}
W^{(2)}_{i_2j}(x,y)=\left(\begin{array}{ccccccc}
\\
\widetilde{W}^{(0)}_{ij}\\
\\
\hline\\
\begin{array}{ccccccc}
0 & 0 & 0 & -iz_1^* & -iz_2^* & iz_1 & iz_2 \\
0 & 0 & \frac{\kappa}{\pi} & 0 & 0 & 0 & 0  \\
0 & - \frac{\kappa}{\pi} & 0 & 0 & 0 & 0 & 0 \\
0 & 0 & 0 & z_1^* & z_2^* & z_1 & z_2 \\
0 & \frac{\kappa}{\pi} \partial^2 & -\frac{\kappa}{\pi} 
\partial^1 & W_{12,1}^{(2)} & W_{12,2}^{(2)} &  W_{12,3}^{(2)} 
& W_{12,4}^{(2)} \\
0 & -\frac{\kappa}{\pi}\partial^2 & \frac{\kappa}{\pi}\partial^1 
& W_{13,1}^{(2)} & W_{13,2}^{(2)} & W_{13,3}^{(2)} & W_{13,4}^{(2)}
\end{array}
\end{array}
\right)\delta^2(x-y),
\end{equation}
with
\begin{eqnarray}
\label{57}
W_{12,k}^{(2)} &=& 2(i\dot z_k^* - z_k^*A^0)  = W_{13,k}^{(2)}, ~~~k=1,2, 
\nonumber \\
W_{12,l}^{(2)} &=& 2(i\dot z_{l-2} - z_{l-2}A^0) = W_{13,l}, ~~~l=3,4,
\end{eqnarray}
and
\begin{equation}
\label{58}
(\alpha_{i_2}^{(2)} )^{\rm T} (x) = ( (\alpha_{i_1}^{(1)})^{\rm T}, 
\alpha_{12}^{(2)}, \alpha_{13}^{(2)}) (x),
\end{equation}
with
\begin{eqnarray}
\label{59}
&&\alpha_{12}^{(2)} = 2(z_a \dot z_a^* - \dot z_a z_a^* ) \dot A^0 
+ 2 |z_a|^2  \partial_m \dot A^m  
+ 4i |z_a|^2 A_\mu \dot A^\mu 
+ 4 \frac{d}{dt} (A^m  z_a \partial_m z_a^*)
\nonumber \\
&&~~~~~  + 2\partial_m A^m \frac{d}{dt} |z_a|^2  
+ 2i  A_\mu A^\mu \frac{d}{dt} |z_a|^2 
- 2i \frac{d}{dt}(z_a\partial_m \partial^m z_a^*)  \nonumber \\
&&\alpha_{13}^{(2)} = 2 (z_a \dot z_a^*  -  z_a^* \dot z_a) \dot A^0 
- 2 |z_a|^2 \partial_m \dot A^m 
+ 4i |z_a|^2 A_\mu \dot A^\mu 
- 4 \frac{d}{dt} (A^m z_a^* \partial_m z_a) 
 \nonumber \\
&&~~~~~ - 2 \partial_m A^m \frac{d}{dt} |z_a|^2 
+ 2i A_\mu A^\mu \frac{d}{dt} |z_a|^2  
- 2i \frac{d}{dt} (z_a^* \partial_m \partial^m z_a). 
\end{eqnarray}
In addition to the previous null eigenvectors, 
we thus have a new null eigenvector
\begin{equation}
\label{60}
\lambda_{i_2}^{(2)}(x,y)= 
(0,0,0,0,0,0,0,0,2\partial_x^1, 2\partial_x^2,0,1,-1) \delta^2(x-y).
\end{equation}
The associated constraint is now found to vanish ``identically"
\begin{equation}
\label{61}
\Omega_1^{(2)} (x) = \int d^2y \lambda_{i_2}^{(2)} (x,y) L_{i_2}^{(2)} (y) 
= 2\partial^1 \alpha_9^{(2)} + 2\partial^2 \alpha_{10}^{(2)} 
+ \alpha_{12}^{(2)} - \alpha_{13}^{(2)} =0.
\end{equation}
The algorithm ends at this point.

The local symmetries of the action (\ref{36}) 
are encoded in the identity (\ref{61}). 
Using the Eqs. (\ref{38}), (\ref{48}) and (\ref{55}), 
the identity (\ref{61}) is equivalent to 
\begin{equation}
\label{62}
\Omega_1^{(2)} = \frac{d}{dt} ( \partial^1 L_2^{(0)} 
+ \partial^2 L_3^{(0)} - iz_1 L_4^{(0)} -iz_2 L_5^{(0)} 
+ i z_1^* L_6^{(0)} + i z_2^* L_7^{(0)} ) + \frac{d^2}{dt^2} L_1^{(0)}
\equiv 0. 
\end{equation}
Comparing this with Eq. (\ref{32}) for the gauge invariant CP$^1$ model, 
we have the following relations
\begin{eqnarray}
\label{63}
\begin{array}{llll}
\phi_1^{2(1)} (x,y) &= \partial_x^1 \delta^2(x-y), 
&~ \phi_1^{3(1)} (x,y) &= \partial_x^2 \delta^2 (x-y) ,\\
\phi_1^{4(1)}(x,y) &= -iz_1 \delta^2 (x-y), &~ \phi_1^{5(1)} (x,y) 
&= -iz_2 \delta^2 (x-y) ,\\
\phi_1^{6(1)}(x,y) &= iz_1^* \delta^2 (x-y), &~\phi_1^{7(1)} (x,y) 
&= iz_2^* \delta^2 (x-y) ,\\
\phi_2^{1(2)}(x,y) &= \delta^2 (x-y),      ~ & ~&~
\end{array}
\end{eqnarray}
while all the others are vanishing. 
Since the action (\ref{36}) is invariant under the transformation
\begin{equation}
\label{64}
\delta \varphi^i (y) = \sum_k \int d^2 x \left( \Lambda_k (x) \phi_k^{i(0)} 
+ \dot \Lambda_k (x) \phi_k^{i(1)} (x,y) + \ddot \Lambda_k (x) \phi_k^{i(2)} 
(x,y) \right ),
\end{equation}
the resulting symmetry transformations in the CP$^1$ model are finally 
obtained as
\begin{eqnarray}
\label{65}
&&\delta A^{\mu} (x) = \partial^{\mu} \Lambda_1 , \nonumber \\
&&\delta z_a (x) = -i z_a \Lambda_1 , \nonumber \\
&&\delta z_a^* (x) = i z_a^* \Lambda_1,
\end{eqnarray}
which are the well--known gauge transformations of the corresponding model. 
As results, we have systematically derived the set of symmetry transformations
starting from the Lagrangian of the gauge invariant CP$^1$ model 
with the CS term.

\vspace{1cm}
\begin{center}
{\section{\bf Conclusion}}
\end{center}

In conclusion, we have considered the Lagrangian approach of
different types of gauge invariant systems, which are the CSM having
the B-type constraints and the CP$^1$ model having the 
mixed A- and B-types ones. 
We have first turned the anomalous CSM into a fully 
first class constrained system, following the BFT method, and 
have shown how the symmetry transformation could be derived on a
purely Lagrangian level, 
especially without resorting to the Hamiltonian formulation.
On the other hand, 
we have also systematically derived the well--known symmetry transformation 
by analyzing CP$^1$ model, which is the different type from the CSM because
this model imposes the CP$^1$ constraint by hand.
As results, we have shown that the Lagrangian approach 
would provide a systematic derivation of symmetry transformation of 
a Lagrangian. 
We hope that this Lagrangian approach employed 
in our derivation 
will be of much interest in complicated Lagrangian whose full local symmetries
can not easily be extracted out, and also
in the context of the field--anti-field formalism
while keeping the manifestly Lorentz covariant partition function.

\vspace{1cm}
\begin{center}
{\section*{Acknowledgements}}
\end{center}

The present study was supported by
the Basic Science Research Institute Program,
Ministry of Education, Project No. BSRI--97--2414.

\newpage
\vspace{1cm}

\begin{center}
\section*{References}
\end{center}

\begin{description}{}
\item{1.} Batalin I A and Fradkin E S 1987 
            {\it Nucl. Phys.} {\bf B279} 514;
            1986 {\it Phys. Lett.} {\bf B180} 157;
\item{}\hspace{0.4cm} Batalin I A and Tyutin I V 1991
           {\it Int. J. Mod. Phys.} {\bf A6} 3255
\item{2.} Banerjee R 1993 {\it Phys. Rev.} {\bf D48} R5467; 
\item{}\hspace{0.4cm} Banerjee R {\it et al.}
            {\it Phys. Rev.} {\bf D49} 5438 
\item{3.} Kim Y--W {\it et al} 1995
          {\it Phys. Rev.} {\bf D51} 2943;
\item{}\hspace{0.4cm} Cha J--H {\it et al.} 1995  {\it Z. Phys.} 
           {\bf C69} 175
\item{4.} Kim W. T. {\it et al.} 1997 {\it J. Phys. G: Nucl. Part.} 
        {\bf 23} 325
\item{5.}  Kim Y--W and Rothe K D 1998 {\it Nucl. Phys.} {\bf B510} 
          511;
\item{}\hspace{0.4cm} Kim Y--W {\it et al.} 1998
          {\it J. Phys. G: Nucl. Part.} {\bf 24} 953
\item{6.} Kim Y--W {\it et al.} 1997
          {\it Int. J. Mod. Phys.} {\bf A12} 4217;
\item{}\hspace{0.4cm} Park M--I and Park Y--J 1998 {\it Int. J. Mod. Phys.} 
          {\bf A13} 2179
\item{7.} Batalin I A and  Vilkovsky G A 1981 {\it Phys. Lett.} 
          {\bf 102B} 27;
          1983 {\it Phys. Rev.} {\bf D28} 2567; 1984 {\it ibid}. Errata 
          {\bf D30} 508; 
\item{}\hspace{0.4cm} Gomis J {\it et al.} 1995 {\it Phys. Rep.} {\bf 259} 1
\item{8.} Sudarshan E C G and Mukunda N 1974 {\it Classical Dynamics: 
          A Modern Perspective} (John Wiley \& Sons)
\item{9.} Shirzad A 1998 {\it J. Phys. A: Math. Gen.} {\bf 31} 2747
\item{10.} Kim Y--W {\it et al.} 1992 {\it Phys. Rev.} {\bf D46} 4574;
\item{}\hspace{0.4cm} Kim S--K {\it et al.} 1995 {\it J. Korean Phys. Soc.} 
                     {\bf 28} 128
\item{11.} Jackiw R and Rajaraman R 1985
            {\it Phys. Rev. Lett.} {\bf 54} 1219;
            1985 {\it Phys. Rev. Lett.} {\bf 54} 2060(E)
\item{}\hspace{0.6cm} Falck N K and Kramer G 1987 
            {\it Ann. Phys.} {\bf 176} 330
\item{}\hspace{0.6cm} Shizuya K 1988 
           {\it Phys. Lett.} {\bf B213} 298
\item{}\hspace{0.6cm} Harada K 1990 
           {\it Phys. Rev. Lett.} {\bf 64} 139
\item{12.}  Kim Y--W and Rothe K D 1998 
           {\it Int. J. Mod. Phys.} {\bf A13} 4183 
\item{13.} Deser S {\it et al.} 1982 {\it Ann. Phys. (N.Y.)}
           {\bf 140} 372
\item{14.} Panigrashi P K {\it et al.}  1988 
           {\it Phys. Rev. Lett.}
           {\bf 61} 2827
\item{15.} Han C 1993 {\it Phys. Rev.} {\bf D47} 5522
\item{16.} Kim Y--W {\it et al.} 1995 {\it J. Korean Phys. Soc.}
           {\bf 28} 773  
\item{17.} Dirac P A M 1964 {\it Lectures on quantum mechanics}
         (Belfer graduate School,
            Yeshiba University Press, New York)
\item{18.} Faddeev L D and Popov V N 1967 {\it Phys. Lett.}
           {\bf B25} 29 
\item{19.} As is well
known, symmetry transformations of the Lagrangian generally imply
restrictions on the symmetry transformations of the total Hamiltonian. 
See also, Henneaux M and Teitelboim C 1992 
{\it Quantization of Gauge Systems}
          (Princeton University Press, Princeton, New Jersey)
\item{20.} Polyakov A M 1988 {\it Mod. Phys. Lett.} {\bf A3} 325
\end{description}
\end{document}